\documentclass[aps,prb,twocolumn,showpacs,superscriptaddress,floatfix]{revtex4-1}
\usepackage{amsmath,amssymb}
\usepackage{graphicx}
\usepackage{epsfig}
\usepackage{color}
\usepackage{rotating}
\usepackage{bm}
\usepackage{setspace}
\begin{document}
\title{Molecular beam epitaxy growth and surface structure of Sr$_{1-x}$Nd$_{x}$CuO$_2$ cuprate films}
\author{Jia-Qi Fan}
\author{Shu-Ze Wang}
\author{Xue-Qing Yu}
\author{Rui-Feng Wang}
\author{Yan-Ling Xiong}
\affiliation{State Key Laboratory of Low-Dimensional Quantum Physics, Department of Physics, Tsinghua University, Beijing 100084, China}
\author{\\Can-Li Song}
\email[]{clsong07@mail.tsinghua.edu.cn}
\author{Xu-Cun Ma}
\email[]{xucunma@mail.tsinghua.edu.cn}
\affiliation{State Key Laboratory of Low-Dimensional Quantum Physics, Department of Physics, Tsinghua University, Beijing 100084, China}
\affiliation{Frontier Science Center for Quantum Information, Beijing 100084, China}
\author{Qi-Kun Xue}
\affiliation{State Key Laboratory of Low-Dimensional Quantum Physics, Department of Physics, Tsinghua University, Beijing 100084, China}
\affiliation{Frontier Science Center for Quantum Information, Beijing 100084, China}
\affiliation{Beijing Academy of Quantum Information Sciences, Beijing 100193, China}

\begin{abstract}
We report epitaxial growth and surface structure of infinite-layer cuprate Sr$_{1-x}$Nd$_{x}$CuO$_{2}$ films on SrTiO$_3$(001) substrates by combining ozone-assisted molecular beam epitaxy and \textit{in-situ} scanning tunneling microscopy. Careful substrate temperature and flux control has been used to achieve single phase, stoichiometric and $c$-axis oriented films. The surface of the films is usually characterized by mixed CuO$_2$ surface and grid-like superstructure. The superstructure exhibits a periodicity of 3.47 nm that corresponds to a coincidence lattice between overlayer peroxide SrO$_2$ and underlying CuO$_2$ plane, and gives rise to conductance spectrum that is distinct from the Mott-Hubbard band structure of CuO$_2$. At higher Nd composition $x$ $>$ 0.1, a $(2\times2)$ surface characteristic of the hole-doped CuO$_2$ emerges, which we ascribe to the intake of apical oxygens in the intervening Sr planes.
\end{abstract}

\maketitle
Infinite-layer (IL) $A$CuO$_2$ ($A$ = Ca, Sr, Ba) compounds exhibit the simplest crystal structure among cuprates, in which the major superconducting CuO$_2$ is alternatively separated by alkaline earth cations along the crystallographic $c$-axis \cite{siegrist1988parent}. Partial substitution of divalent $A^{2+}$ ions by trivalent ions such as La$^{3+}$ and Nd$^{3+}$ leads to electron-doped superconductivity with a record transition temperature $T_\textrm{c}$ of 43 K \cite{smith1991electron, chen2002Strongly, armitage2010progress}. More remarkably, IL compounds represent a rare category of cuprate superconductors with surface termination of the superconducting CuO$_2$ planes \cite{koguchi1995atomic, Harter2015doping, Zhong2018atomic}. Given that most cuprates are terminated with non-CuO$_2$ charge reservoir layers upon cleaving, e.g.\ BiO for bismuth-based cuprates, this peculiar feature provides an unprecedented opportunity to directly characterize the superconducting CuO$_2$ planes by surface-sensitive experiments \cite{Zhong2019direct}, compared to previous studies \cite{Damascelli2003angle, fischer2007scanning, ye2013visualizing}. A systematic direct measurement of the major CuO$_2$ planes may help understand eventually the microscopic mechanism of high-$T_\textrm{c}$ superconductivity\cite{chen2002Strongly, Zhong2019direct, Harter2012nodeless, misra2002atomic, Lv2015mapping, lv2016electronic, zhong2016nodeless}. However, IL cuprates with tetragonal structure are thermodynamically unstable. It is nearly impossible to synthesize single crystals by conventional solid state methods, and only some powder form of IL samples was obtained using high pressure techniques\cite{takano1989acuo2, er1994high}.

Epitaxial films of IL cuprates can be stabilized and prepared on appropriate substrates by using pulsed laser deposition (PLD)\cite{gupta1994thin, leca2006superconducting, Tomaschko2012properties} or reactive molecular beam epitaxy (MBE) technique\cite{karimoto2001single, krockenberger2011growth, krockenberger2012molecular, Ikeda2019molecular}. However, the as-grown thin films are often characterized with several competing phases, such as Sr$_2$CuO$_3$, Sr$_{14}$Cu$_{24}$O$_{41}$ and orthorhombic SrCuO$_2$ \cite{krockenberger2018infinite}, as summarized in Table I. Furthermore, due to the limited solubility of trivalent ions in IL compounds, oxygen-deficient or -redundant superstructures with a relatively larger out-of-plane lattice parameter, referred as a long-\textit{c} phase, occur at elevated doping \cite{Zhong2019direct, leca2006superconducting, karimoto2001single, gupta1994thin}. In this study, we combine ozone-assisted MBE and \textit{in-situ} scanning tunneling microscopy (STM) to solve these problems, aiming to establish growth procedures for single phase crystalline Sr$_{1-x}$Nd$_{x}$CuO$_{2}$ (SNCO, 0.08 $\leq x \leq$ 0.12) thin films. We emphasize that, compared to alternative shutter-controlled deposition, our method for composition/phase control is self-regulated, without the complicated calibration of the composition by shutter time.

The experiments were performed on a commercial ultrahigh vacuum (UHV) STM apparatus (Unisoku), connected to an ozone-assisted MBE chamber for \textit{in-situ} film growth. Nb-doped SrTiO$_3$(001) substrates were firstly degassed at 600$^\textrm{o}$C, and subsequently annealed at 1250$^\textrm{o}$C under UHV for 20 minutes to get the clean surface. Prior to film epitaxy, fluxes of all metal sources (Sr, Nd and Cu) were precisely calibrated in sequence by using a standard crystal microbalance (QCM, Inficon SQM160H). Epitaxial thin films were then prepared by co-deposition of high-purity metal sources from standard Knudsen cells under an ozone flux beam of $\sim$ 1.1 $\times$ 10$^{-5}$ Torr. The growth rate is 0.4 unit cell per minute, and the flux ratio between Nd and Cu sources is used to calculate the nominal composition $x$. Polycrystalline PtIr tips were cleaned by electron-beam heating and calibrated on MBE-grown Ag/Si(111) films. Tunneling spectra were measured using a standard lock-in technique with a small bias modulation of 10 mV at 937 Hz. After \textit{in-situ} STM measurements at 78 K, the samples were taken out from the UHV chamber for X-ray diffraction (XRD) measurements using the monochromatic Cu K$_{\alpha1}$ radiation with a wavelength of 1.5406 \AA.

\begingroup
\begin{table}[h]
\center
\caption{\label{tab:crystal}Crystal structure and lattice parameters for Sr-Cu-O compounds in the thermodynamic proximity of IL cuprates.}
\begin{footnotesize}
\scalebox{0.90}{
\begin{tabular}{ l c c c c r }
  \hline\hline
 & Space group & \textit{a}(${\textrm{\AA}}$) & \textit{b}(${\textrm{\AA}}$) & \textit{c}(${\textrm{\AA}}$) & Ref  \\
  \hline
   IL tetragonal SrCuO$_{2}$ & \textit{P}4/\textit{mmm}& 3.9269 & = \textit{a} & 3.4346 & \cite{siegrist1988parent}  \\
   Orthorhombic SrCuO$_{2}$ & \textit{Cmcm}& 3.5770 & 16.342 & 3.9182 & \cite{matsushita1995growth} \\
   Orthorhombic Sr$_{2}$CuO$_3$ & \textit{Immm}& 12.702 & 3.911 & 3.4990 & \cite{Hyatt2004high} \\
   Orthorhombic Sr$_{14}$Cu$_{24}$O$_{41}$  & \textit{Amma}& 11.488 & 13.414 & 27.428 & \cite{abbamonte2004crystallization} \\
   Tetragonal SrO$_2$  & \textit{I}4/\textit{mmm}& 3.55 & = \textit{a} & 6.55 & \cite{middleburgh2013accommodation} \\
  \hline
\end{tabular}}
\end{footnotesize}
\end{table}
\endgroup

\begin{figure*}[t]
\includegraphics[width=1.97\columnwidth]{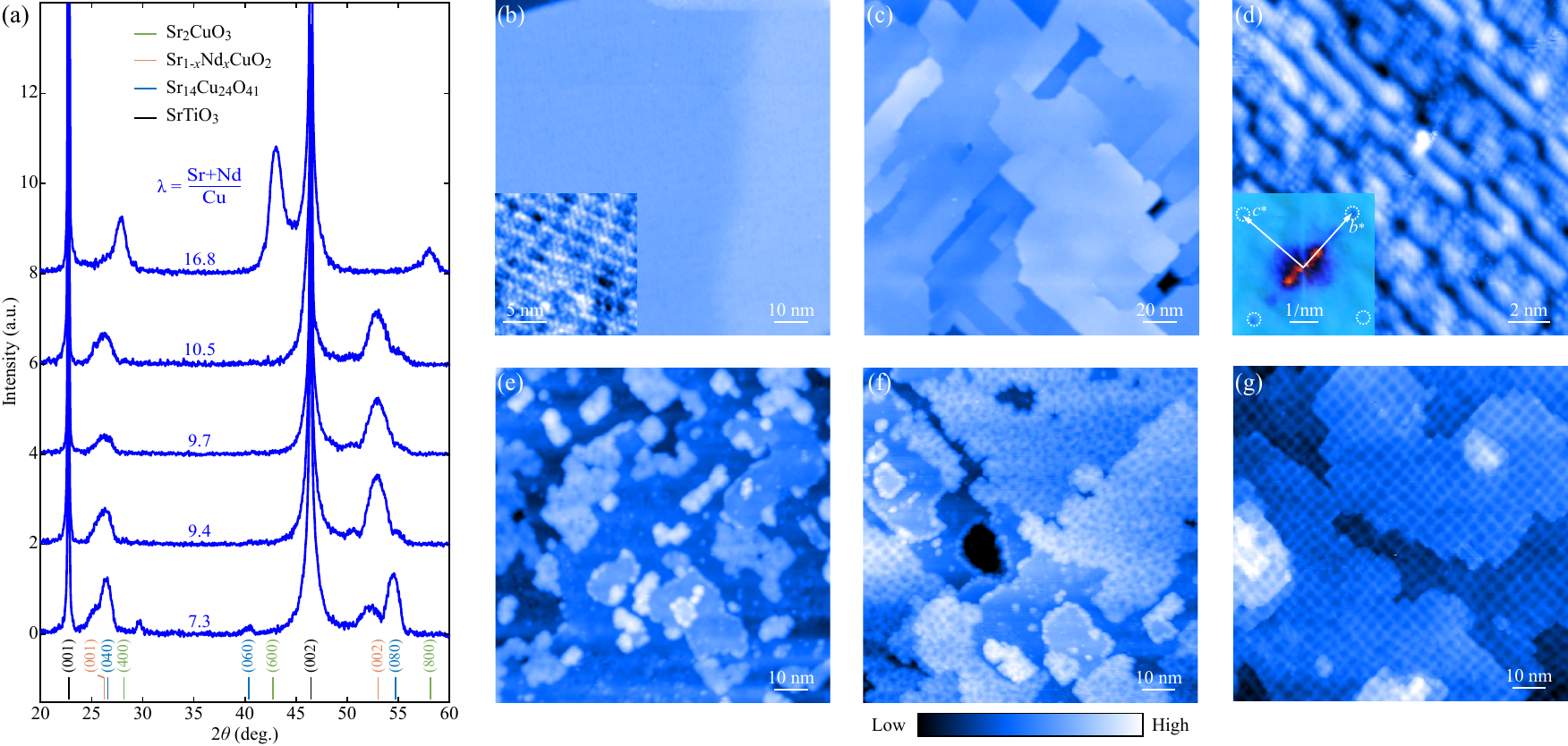}
\caption{(color online) (a) Representative XRD patterns of epitaxial films grown with various flux ratio (Sr+Nd)/Cu as indicated. The color vertical bars correspond to the indexation of the crystal structure database for different phases referred. (b) STM topography (100 nm $\times$ 100 nm, $V$ = $-$5.5 V, $I$ = 20 pA) of spin ladder Sr$_{14}$Cu$_{24}$O$_{41}$ at a small $\lambda$ of 7.3. Inserted is a zoom-in STM image of the chain-like (010) surface (20 nm $\times$ 20 nm, $V$ = $-$4.0 V, $I$ = 20 pA). (c) Large-scale STM topography (200 nm $\times$ 200 nm, $V$ = $-$4.0 V, $I$ = 20 pA) of Sr$_2$CuO$_3$ at a large $\lambda$ of 16.8. (d) Atomic-resolved STM image of Sr$_2$CuO$_3$ (16 nm $\times$ 16 nm, $V$ = $-$4.5 V, $I$ = 15 pA). Inset shows the corresponding FFT image, with $b^*$ and $c*$ denoting the two reciprocal lattice vectors. (e-g) Morphographies (100 nm $\times$ 100 nm, $I$ = 20 pA) of IL SNCO cuprate films with increasing $\lambda$. The sample bias $V$ for STM imaging is (e) 3.0 V, (f) $-$4.0 V and (g) $-$3.5 V. The Nd composition $x$ is 0.08 in (e, f) and 0.10 in (g).
}
\end{figure*}

Growth of IL SNCO epitaxial films demands for precise control of the substrate temperature $T_{\textrm{sub}}$ and cation stoichiometry. Similar to previous reports \cite{mihailescu2014origin}, we found that tetragonal IL SNCO films start to crystallize at 500$^\textrm{o}\textrm{C}$ and change to orthorhombic phase above 610$^\textrm{o}\textrm{C}$. Thus, $T_{\textrm{sub}}$ = 550$^\textrm{o}\textrm{C}$ was chosen for both good crystallinity and avoiding high temperature orthorhombic phase. Figure 1(a) shows the XRD patterns of as-grown films as a function of the nominal flux ratio $\lambda$ = (Sr+Nd)/Cu, with a smaller Nd/Cu flux ratio of $x \leq$ 0.10. Apparently, IL SNCO phase coexists with Sr-deficient spin ladder Sr$_{14}$Cu$_{24}$O$_{41}$ at lower $\lambda$ of 7.3. This is understandable because Sr has a higher vapor pressure of 1.8 $\times$ 10$^{-2}$ Torr and is very volatile at $T_{\textrm{sub}}$ = 550$^\textrm{o}\textrm{C}$. Meanwhile, Sr is easily oxidized in ozone atmosphere, which reduces its effective flux during the growth. The two factors explain why a larger $\lambda$ $\geq$ 9.4 is required to prepare single phase IL films, as demonstrated by the XRD spectra in Fig.\ 1(a). Evidently, the cation stoichiometry of SNCO is quasi-self-regulating, resembling, to some extent,  the growth of GaAs and metal chalcogenides \cite{li2010intrinsic, Song2011molecular}. We note that the self-regulation of stoichiometry is somewhat limited and the IL SNCO phase forms only in a narrow window of $\lambda$. A larger $\lambda$ of 16.8 converts the epitaxial films to a more thermodynamically stable Sr$_2$CuO$_3$ phase [see Fig.\ 1(a)].

Our STM characterization corroborates the flux-ratio-dependent phase evolution. At $\lambda$ = 7.3, the chain-like surface characteristic of spin ladder Sr$_{14}$Cu$_{24}$O$_{41}$(010) occurs [Fig.\ 1(b)], whereas single phase Sr$_2$CuO$_3$ overwhelms the others under Sr-rich condition [Figs.\ 1(c) and 1(d)]. Fast Fourier transform (FFT) analysis inserted in Fig.\ 1(d) indicates that the in-plane lattice constants are \textit{b} = 3.9 $\pm$ 0.1 $\textrm{\AA}$ and \textit{c} = 3.5 $\pm$ 0.1 $\textrm{\AA}$, consistent with the expected value for orthorhombic Sr$_2$CuO$_3$(100) surface in Table I. The single phase IL SNCO films are prepared at an intermediate $\lambda$ and display atomically flat surface [Figs.\ 1(e)-1(g)], which are separated by gird-like superstructure. The grid-like feature gradually becomes prominent with increasing $\lambda$ and covers the whole surface at $\lambda$ $\sim$ 10.5.

\begin{figure}[h]
\includegraphics[width=\columnwidth]{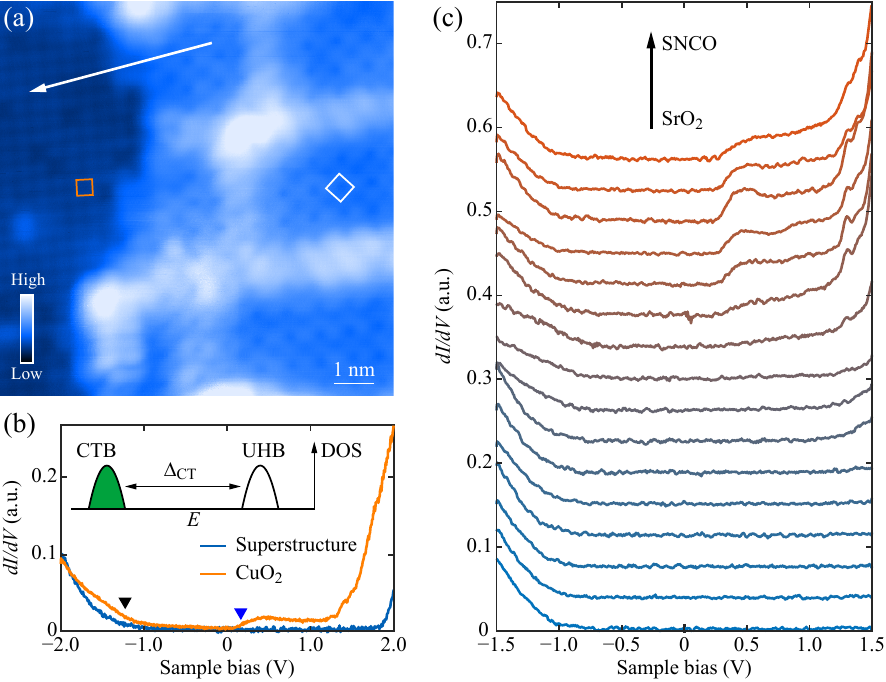}
\caption{(color online) (a) Atomically resolved topography (10 nm $\times$ 10 nm, $V$ = $-$2.0 V, $I$ = 20 pA) across a step edge separating CuO$_2$ plane (left side) and grid-like superstructure (right side) in Sr$_{0.9}$Nd$_{0.1}$CuO$_{2}$. Orange and white squares denote the respective in-plane unit cells. (b) Spatially-averaged tunneling spectra on CuO$_2$ and grid-like superstructure. Inserted is the schematic band structure of pristine cuprates displaying the UHB (unfilled) and CTB (green). The black and blue triangles mark the onsets of CTB and UHB throughout. Setpoint: $V$ = $-$2.0 V, $I$ = 100 pA. (c) A series of \textit{dI/dV} spectra acquired along the white arrow in (a). Setpoint: $V$ = $-$1.5 V, $I$ = 20 pA.
}
\end{figure}

To identify the two apparently distinct surfaces of IL SNCO films, we acquire atomically-resolved STM images, as illustrated in Fig.\ 2(a). The flat surface has a square lattice with a periodicity of $\sim$ 3.9 \AA, matching well CuO$_2$-terminated IL SNCO \cite{smith1991electron, Zhong2019direct}. This is indeed supported by the site-dependent differential conductance \textit{dI/dV} spectra in Fig.\ 2(b). On the flat surface, the tunneling \textit{dI/dV} spectrum features a fundamental Mott-Hubbard band structure of the cuprate CuO$_2$ planes, accompanied by metallic-like states within the charge-transfer gap \cite{Zhong2019direct}. It is worth noting that the Fermi level $E_F$ is closer to the upper Hubband (UHB) than the charge-transfer band (CTB), in line with the electron doping by the Nd$^{3+}$ substitution for Sr$^{2+}$ ions.

\begin{figure*}[t]
\includegraphics[width=1.937\columnwidth]{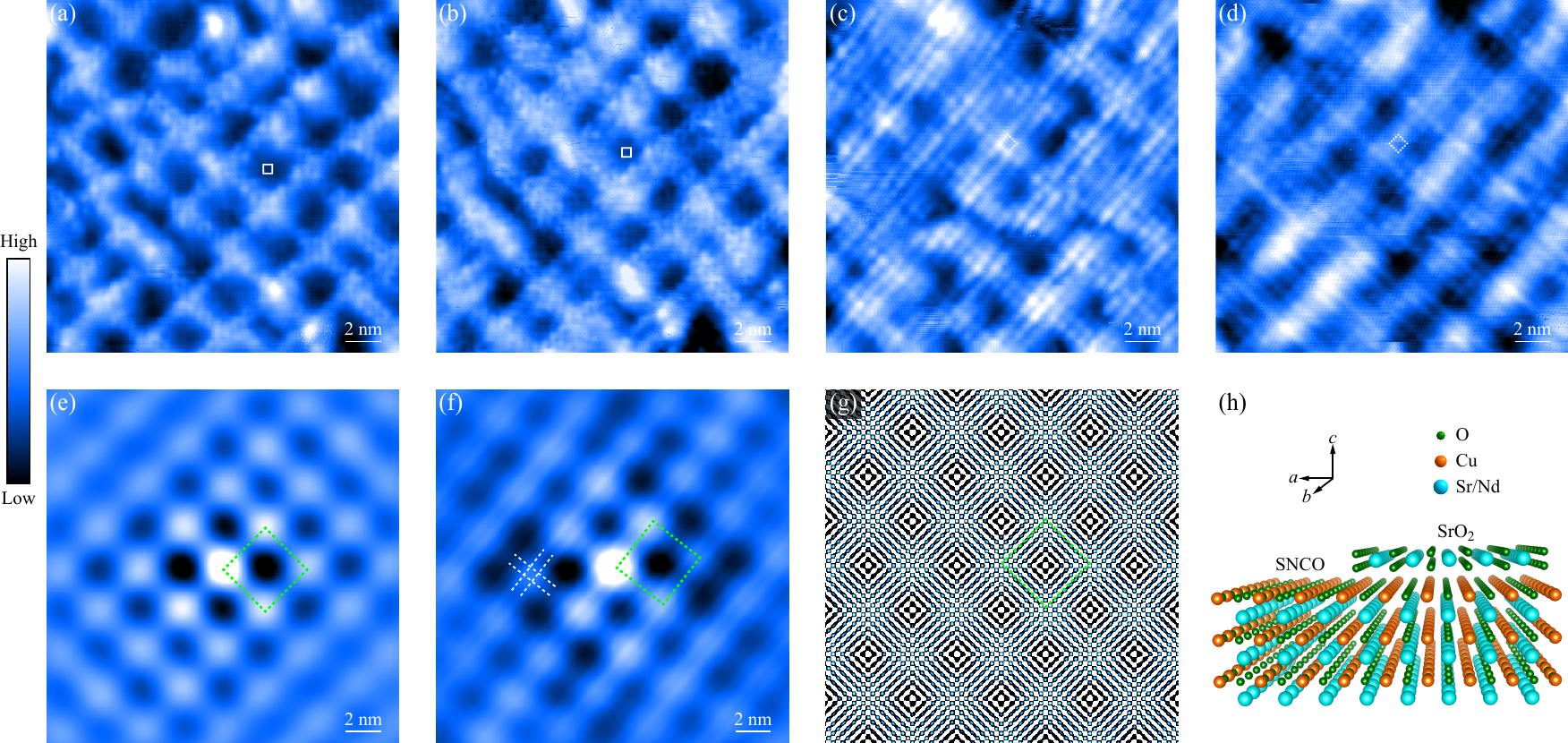}
\caption{(color online) (a-d) Bias-dependent STM images (20 nm $\times$ 20 nm, $I$ = 20 pA) of the coincidence lattice between SrO$_2$ overlayer and CuO$_2$ plane. The bias $V$ is (a) $-$3.0 V, (b) $-$2.0 V, (c) 3.0 V and (d) 4.0 V. Note that the surface structure alters from SrO$_2$(001)$-$$\sqrt{2} \times \sqrt{2}$ R45$^\textrm{o}$ (solid squares) to SrO$_2$(001)$-$$2 \times 2$ (dashed squares) as the bias polarity is reversed. (e, f) Autocorrelation analysis of the STM images in (a) and (c), respectively. The green squares represent the unit cells of grid-like superstructure. (g) Simulated Moir\'{e} pattern between SrO$_2$ and CuO$_2$. (h) Schematic sketch of the SrO$_2$ overlayer on CuO$_2$-terminated SNCO films.
}
\end{figure*}

In contrast, the grid-like superstructure is characterized by a larger in-plane unit cell of $\sim$ 5.0 $\textrm{\AA}$ (marked by the white square), rotated by 45$^\textrm{o}$ relative to the CuO$_2$ unit cell in Fig.\ 2(a). A possible surface reconstruction of SNCO(001)$-$$\sqrt{2}\times\sqrt{2}$ R45$^\textrm{o}$ could be safely excluded since the measured periodicity of $\sim$ 5.0 $\textrm{\AA}$ deviates substantially from the $\sqrt{2}$ times ($\sim$ 5.6 $\textrm{\AA}$) of in-plane lattice constant of SNCO. Moreover, tunneling \textit{dI/dV} spectrum of gird-like superstructure shows an extremely large band gap ($\sim$ 2.8 eV) and is significantly different from that of CuO$_2$ plane [Fig.\ 2(b)]. This is confirmed by the linecut \textit{dI/dV} spectra across one step edge between the grid-like superstructure and the CuO$_2$ surface in Fig.\ 2(c). These observations, together with the populated gird-like superstructure at elevated $\lambda$ [Figs.\ 1(e)-1(g)], strongly suggest that the superstructure originates from a totally different compound, most probably linking with strontium. Tetragonal strontium peroxide SrO$_2$ has a lattice constant of 3.55 $\textrm{\AA}$ in the \textit{a-b} plane (Table I) \cite{middleburgh2013accommodation}, coinciding with 1/$\sqrt{2}$ of the measured unit cell periodicity of 5.0 $\textrm{\AA}$ in Fig.\ 2(a). In other words, the grid-like surface might correspond to SrO$_2$ in nature, which exhibits an enlarged surface structure, i.e.\ SrO$_2$(001)$-$$\sqrt{2}\times\sqrt{2}$ R45$^\textrm{o}$. Considering that no excess phase other than IL SNCO is found in the bulk-sensitive XRD spectra at intermediate $\lambda$ [Fig.\ 1(a)], the SrO$_2$ ought to occur only at the topmost CuO$_2$ surface of epitaxial SNCO films.

\begin{figure}[h]
\includegraphics[width=\columnwidth]{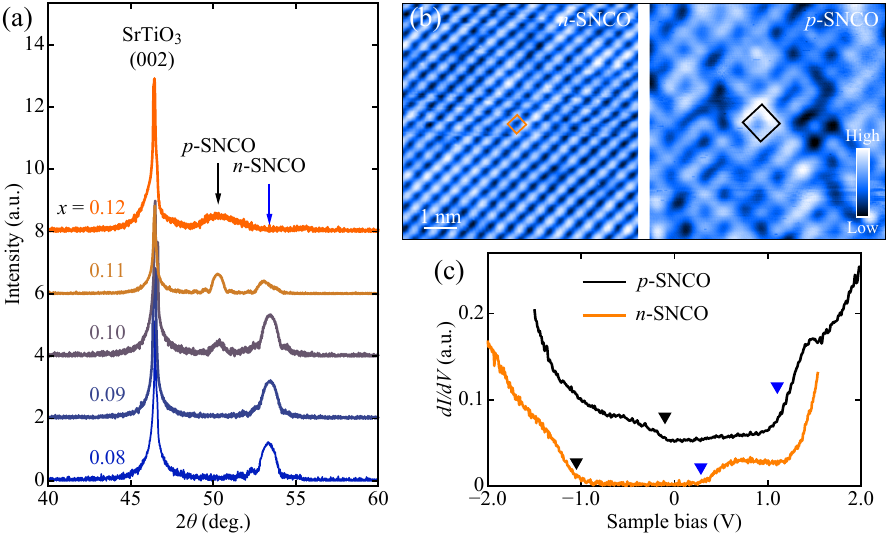}
\caption{(color online) (a) XRD spectra of IL SNCO films with varying $x$. Blue and black arrows denote the reflection
peaks from $n$- and $p$-SNCO, respectively. (b) STM topographies (7 nm $\times$ 7 nm, $I$ = 20 pA) of coexisting $n$-SNCO (left panel, $V$ = $-$1.5 V) and $p$-SNCO films (right panel, $V$ = $-$3.0 V) at $x$ = 0.12. The unit cells outlined by colored squares become doubled in size for $p$-SNCO as compared with $n$-SNCO. (c) Comparison between tunneling \textit{dI/dV} spectra on $n$-SNCO ($V$ = $-$2.0 V, $I$ = 100 pA) and $p$-SNCO films ($V$ = $-$1.5 V, $I$ = 100 pA).
}
\end{figure}

By acquiring bias-dependent STM images with atomic-scale resolution in Figs.\ 3(a)-3(d), we further confirm this conclusion for the grid-like superstructure. Intriguingly, the SrO$_2$(001)$-$$\sqrt{2}\times\sqrt{2}$ R45$^\textrm{o}$ surface switches to SrO$_2$(001)$-$$2\times2$ structure as the bias polarity is reversed from negative to positive. This hints that the emergent surface structures, irrespective of $\sqrt{2}\times\sqrt{2}$ and $2\times2$, may most likely stem from charger ordering in SrO$_2$ \cite{renner2002atomic, Iwaya2011stripe}. The surface structure switching should be due to a bias-dependent lateral variation of local density of states in SrO$_2$ \cite{Maroutian2003superstructure}, which requires further theoretical investigations. Notwithstanding, the grid-like superstructure remains unchanged in both dimension and orientation. The measured periodicity is 34.7 $\pm$ 1.4 $\textrm{\AA}$ on average, which is approximately 10 times the Sr-Sr atom spacing ($a_{\textrm{SrO$_2$}}$ $\sim$ 3.55 $\textrm{\AA}$) in SrO$_2$ according to the autocorrelation analysis in Figs.\ 3(e) and 3(f). Additionally, the possible $2\times2$ charge ordering of SrO$_2$ is apparently visible (see the white dashes) in Fig.\ 3(f) that enables to deduce the zero angle of intersection between the respective lattices of SrO$_2$ and grid-like superstructure. Note that the latter periodicity of 34.7 $\pm$ 1.4 $\textrm{\AA}$ coincides nicely with 9 times of the lattice constant $a_\textrm{SNCO}$ of SNCO films \cite{smith1991electron, bobrovskii1997neutron}, a coincidence lattice between the SrO$_2$ overlayer and CuO$_2$-terminated SNCO films is proposed to be responsible for the grid-like superstructure [Figs.\ 3(g) and 3(h)]. Figure 3(g) illustrates a simulated Moir\'{e} pattern by reasonably assuming $a_{\textrm{SrO$_2$}}$ = 3.55 $\textrm{\AA}$ and $a_\textrm{SNCO}$ = 3.94 $\textrm{\AA}$, which matches well our results [Figs.\ 3(e) and 3(f)].

The coincidence lattice for the superstructure, rather than a simple topographic Moir\'{e} pattern between the SrO$_2$ overlayer and underlying CuO$_2$, is based on two experimental findings. One is the significant dependence of the apparent corrugation of grid-like superstructure on the applied sample voltage in Figs. 3(a)-3(d). For example, the corrugation of superstructure is more apparent at negative biases. The other finding relates to the local distortion in the grid-like superstructure and the accompanying charge ordering, which is unexpected for Moir\'{e} pattern. Instead, it can be the local structural distortion in coincidence lattice to yield the bias-dependent corrugation, distorted superstructure and charge ordering.

Next we explore the dependence of SNCO films on the nominal composition $x$ of Nd. As shown in Fig.\ 4(a) are five XRD spectra of IL SNCO films at varied $x$. Analogous to La-doped Sr$_{1-x}$La$_{x}$CuO$_2$ (SLCO) IL epitaxial films \cite{Zhong2019direct}, a second phase with a larger $c$-axis lattice constant emerges at $x >$ 0.1, coexists and becomes dominant with increasing $x$. The emergent new phase is characteristic of CuO$_2$(001)$-2\times2$ surface structure [Fig.\ 4(b)] and exhibits a hole-doped behavior with the $E_\textrm{F}$ closer to CTB [see the black curve in Fig.\ 4(c)], which we dub as $p$-SNCO. In contrast, the electron doped $n$-SNCO films always display a bare CuO$_2$(001)$-1\times1$ surface, even in the two-phase coexisting SNCO films for $x$ = 0.12 [Figs.\ 4(b) and 4(c)]. Without loss of generality, we attribute the CuO$_2$(001)$-2\times2$ surface reconstruction and emergent $p$-type behavior in SNCO films as the considerable incorporation of apical oxygens in the intervening Sr planes\cite{Zhong2019direct}, which overwhelm the electron doping by Nd$^{3+}$ donors. In any case, the observed tunneling \textit{dI/dV} spectra are of striking resemblance, except for an energy shift in $E_\textrm{F}$. This echoes the self-modulation doping scheme\cite{Zhong2019direct}, namely doping the intervening Sr layers changes the fundamental Mott-Hubbard band structure of CuO$_2$(001) little.

Finally we comment on implication from the observed SrO$_2$ overlayers. Based on the step height in Fig.\ 2(a), we readily estimate the thickness of SrO$_2$ overlayer, to wit, only half of unit cell ($\sim$ 3.3 $\textrm{\AA}$). Evidently, the top SrO$_2$ layer is insulating and exhibits a large semiconducting gap of $\sim$ 2.8 eV [Fig.\ 2(b)]. Notably, the surface stacking of one SrO$_2$ layer on CuO$_2$ is structurally similar to the BiO-terminated Bi$_2$Sr$_2$CaCu$_2$O$_{8+\delta}$ \cite{Damascelli2003angle, fischer2007scanning, misra2002atomic}, i.e.\ insulating Sr(Bi) oxides on CuO$_2$. Here, the measured \textit{dI/dV} spectra appear sharply different between SrO$_2$ and CuO$_2$, and thus how the cuprate database from the vacuum-cleaved BiO planes represents the spectral properties of buried CuO$_2$ merits further investigations.

\begin{acknowledgments}
The work is financially supported by the Ministry of Science and Technology of China (Grants No.\ 2017YFA0304600, No.\ 2018YFA0305603), the National Natural Science Foundation of China (Grants No.\ 11774192, No.\ 11634007), and in part by Beijing Innovation Center for Future Chips, Tsinghua University.
\end{acknowledgments}

%\bibliography{SNCO}
%

\end{document}